\documentclass[11pt]{article}
\usepackage{epsfig}
\usepackage{verbatim}
\usepackage{amsmath}
\usepackage{wasysym}
\usepackage{makeidx,showidx}
\usepackage{pst-node}  
\usepackage[utf8]{inputenc} 
\usepackage[pdfmenubar=true, letterpaper=false, pdfpagelabels=true]{hyperref}

\begin{document}

\def\mvec#1{{\bm{#1}}}   

\title{Debunking some myths about biometric authentication}

\author{Alfredo Esposito \\
InfoCert S.p.A.\footnote{Disclaimer: The views expressed in this paper are solely the
author's and should not be attributed to his employer }\\
{\small (alfredo.esposito@infocert.it)}
}

\date{}  %

\maketitle

\begin{abstract}
Biometric authentication systems are presented as
the best way to reach high security levels
in controlling access to IT systems or sensitive infrastructures. But several issues are often not taken properly into account.
In order for the implementation of those systems to be successful, the hidden risks and the related liabilities have to be carefully analyzed before biometrics can be used on a large scale for sensitive applications.
\end{abstract}

\vspace{2.0cm}  

\section{Introduction}
Access control to Information Technology (IT) systems or to sensitive infrastructure is a 
basic tool of security management, keeping the bad guys away from vital 
 data, bank accounts or top secret military projects. Till few years ago access control was based 
almost exclusively on some secrets, typically a password, shared by the person willing to access and the system to be accessed. But passwords can be guessed or, being hard to remember, written down on a piece of paper, easy to get stolen or lost.

With the flourishing of online services and the related boom of frauds and identity thefts a more robust access control system has become necessary. For a while the new paradigm was the so called strong (or two-way) authentication: something to know and something to own. In order to access the user needs to know a personal code and to keep some physical token under his/her sole control, a one-time-password generator or a smart card or a USB token protected by a Personal Identification Number (PIN).
Surely more secure, since losing the token is almost harmless being it useless without the corresponding access code (PIN). Nevertheless it obliges the loser to stay for a while without access capability and to enroll for a new token. 
So authentication based on biometric systems (the use of a physical or behavioural personal feature) has gained momentum, pushed by smart marketing slogans: authentication based on a personal feature that can never be lost, stolen or forgotten. Therefore biometrics is now sold as the silver bullet, the ultimate solution to overcome the annoying aspects of the authentication process. 

But is biometrics really so effective? Are the hard facts supporting the promises of biometrics? 

In the following we will try, without any claim of thoroughness,\footnote{See \cite{NSF} for a more complete treatment.} to look at some often neglected\footnote{In fact most issues addressed in the present paper 
have been analyzed and taken into account in the biometric scientific literature 
(see \cite{NSF} and chapter 1 of  \cite{Mantoni}) but overlooked in the documents addressed to the managers in charge of taking decision, obviously non specialist readers.} aspects which suggest some cautions when evaluating an access control system based on biometric techniques. 
 
\section{Access Control: Some basic concepts}
Let us start with some definitions. 
The basic concepts are: {\it authentication} and {\it identification}. 
\begin{description}
\item[authentication:] the process of checking the validity of an identity claim by
                       matching a credential against a set of reference values;\footnote{Often called {\it verification} in biometric literature.}
\item[identification:] the process of searching the entire set of possible identities 
                       in order to find the right one matching the measured feature.
\end{description} 
More formally, given a reference set $R$ of pairs ({\it user-id}, {\it secret)} 
$$R = \{(u_{1},s_{1}), (u_{2},s_{2}), ...(u_{n},s_{n})\}\,,$$  
the authentication process challenges the user claiming to be ${\it u_{i}}$ to enter the corresponding secret ${\it s_{i}}$. Then the secret ${\it s'}$ entered by the would-be user is compared 
to the corresponding stored value. If there is a match ($s_{i}=s')$, the claim is accepted and 
access is allowed, otherwise it is denied. Pretty deterministic.\footnote{In most cases only a cryptographically hashed digest of the secret is stored on the server site, so leaving an extremely small chance of secret collisions (for an introduction see \cite{Crypt}).}

Identification is a bit more complicated. Again, given a stored set of pairs ({\it user-id, distinguishing feature}), the identification process looks up in the reference set if there is
a distinguishing feature\footnote{Obviously assuming that the feature (or a set of them) is unique.} matching the presented one. A typical forensic example is the usage of fingerprints in order to identify the people involved in a crime. Identification is not used in the traditional systems and it is considered one of the best advantages of biometrics, allowing  the credentials to be presented by the user automatically, sometimes unwillingly, e.g. by using face pictures taken by surveillance cameras in crowded areas.

Another peculiarity of biometrics is the so called {\it negative identification} aiming to prove that a person, 
for example trying to get some benefits of the social security system using a fake identity, is not who he/she claims to be. Also this feature is not available in the traditional systems. 

The caveats discussed in next sections apply to all kinds of  biometric applications.  
\section{Biometrics is different}
There is a number of important aspects that makes biometrics different from the traditional techniques but the most relevant is surely its probabilistic nature.

In fact, several factors contribute to make biometrics not deterministic:
\begin{itemize}  
  \item environmental noise when biometric data are collected, both at the enrollment and during the normal operation; 
  \item context of data acquisition: in a laboratory or into the wild;
  \item cooperative (or not) approach by the end users;
  \item natural changes in the biometric features:
  \begin{itemize}

    \item faces change aging
    \item fingerprints can be worn or modified by age, hard manual works or accidents;
    \item voice changes with age and health status, e.g. if a person is running a cold;
    \item handwritten signature changes.
  \end {itemize}
\end {itemize}
All these changes occur in a way substantially unpredictable; so a system based on them shall cope with results that are different each time. In order to do it the system shall compute a similarity 
score $S$ between the stored reference value and the measured one. Then that score is compared with a predefined threshold $T$ in order to decide whether accept (when \mbox{$S > T$}) or reject the user credentials. The higher the score, more likely the measured sample comes from the right person.
Moreover, given the natural evolution of the features, the system shall also foresee a cyclic renewal 
of the reference sample against which the comparison is made. 

The aleatory nature of biometric data changes the way we evaluate if such a system is suitable for our purposes. In other words, if it fulfils the required specific discriminatory power.
To deal properly with that issue we need to recall a bit of theory of probability. 

In fact, some misunderstandings\footnote{Sometimes also skilled scientists make mistakes! \cite{BadPhys}} about some basic concepts of such theory can lead to totally wrong evaluation of the real effectiveness of the device under study. 
\subsection{False match rate and false non match rate}\label{ss:match}
The more `serious' suppliers of biometric terminals, trying
to underpin the amazing statements about the performances of their 
products, highlight two typical parameters to characterize
the devices: {\it false match rate} (\mbox{FMR}) and {\it false
non match rate} (\mbox{FNMR}).\footnote{\mbox{FMR} and \mbox{FNMR} are both functions of the threshold $T$. 
Generally speaking, \mbox{FMR($T$)} increases lowering $T$, that is making the system more tolerant, while, on the contrary, \mbox{FNMR($T$)} increases making the system more restrictive, i.e. increasing $T$. So a trade off between the two parameters has to be found. Another parameter used quite often is the {\it Equal Error Rate} \mbox{EER} that is the value at the threshold such that \mbox{FMR($T$)=FNMR($T$)}. Other parameters are available and can be important. If, for example, the system can be tuned in order to adjust \mbox{FMR} and \mbox{FNMR} to the system's scope, the {\it Receiver Operation Curve} \mbox{ROC}, plotting \mbox{FMR} vs \mbox{(1-FNMR)}, becomes relevant. Although important when implementing a real system, \mbox{ROC} and other available parameters will not be addressed here. 
This choice does not affect the conclusions of this paper.} 
\begin{itemize}
\item \mbox{FMR} is defined as the probability that there is a match between the reference credential and a false presented one; in other words it is {\it the probability that the device allows the access} {\bf if the user is a fake};
\item \mbox{FNMR} is defined as the probability that there is no match between the reference credential and the true presented one; in other words it is {\it the probability that the device denies the access} {\bf if the user is a legitimate one}. 
\end{itemize}

In the theory of probability these are named conditional probabilities, the condition being in bold.\footnote{In fact EVERY probability is conditional to the application context and the knowledge status of the person assigning it a value. Even the probability $ \frac{1}{2} $ to obtain a head tossing the classical coin implies the unspoken hypotheses that the coin is fair and  that we can ignore the probability of the coin stands on its edge. The conditional probability  to events and facts not strictly related to the problem under examination is called {\it a priori}. For a relatively short but complete introduction see Ref. \cite{Colombo}. \label{note:probcond}}

However, {\it neither} \mbox{FMR} {\it or} \mbox{FNMR} 
{\it correspond to what we are really interested in}, that is the probability that the user is legitimate (or a fake) {\bf if the credential is accepted}. To get this number we shall perform a so called probabilistic inversion that, to avoid trivial mistakes, 
has to be done according to the rules of the probability theory. As we will see with a simple numerical example, $ \mbox{FMR}=0.001 $ DOES NOT MEAN that the probability that a fake is allowed to access the system is 1\permil. 
There are in fact other `ingredients' to be taken into account and we need now to take
a detour in the theory of probability.

\section{The probabilistic inversion and how it
depends on the `sample population'}\label{sec:probinv} 
Let us start introducing a bit of notation. 

Be $A$ the event {\it access allowed} and $F$ the event {\it fake credential}.
Be $\overline {A} $ and $\overline {F}$ the logical complements: {\it access not allowed} and {\it valid credential}, respectively. With this definition our \mbox{FMR} and \mbox{FNMR} become:
\begin{description}
\item[\mbox{FMR}:] $P(A\,|\,F)$ 
\item[\mbox{FNMR}:]  $P(\overline{A}\,|\,\overline{F})$\,
\end{description}
In fact, the right expression should be $P(A\,|\,F,I)$ where $I$ indicates the knowledge status about `the rest of the world' (see note \ref{note:probcond}); nonetheless, for the sake of simplifying the notation, $I$ will not stated explicitly in the following.\\
As said before, these two probabilities should not be confused, as it is often the case (see e.g. \cite{Colombo}), with what we are ultimately interested in:
\begin{description}
\item[{\it probability of 
allowing legitimate access} (\mbox{LA}):] $P(\overline{F}\,|\,A)$
\item[{\it probability of 
denying illegitimate access} (\mbox{NLA}):] $P(F\,|\,\overline{A})$\,.
\end{description}
It is then crucial to learn how \mbox{LA} and \mbox{NLA} are related to \mbox{FMR} and \mbox{FNMR}. We are talking about {\it probability inversion} and the tool to perform this logical operation in the framework of
probability theory is called {\it Bayes' theorem}. 

\subsection{Conditional probabilities and Bayes' theorem}
\label{ss:Bayes}
Let us remind this very simple result of probability theory. 
Given two generic events 
$E$ and $C$, where $C$ is a possible cause for $E$, 
the conditional probability $P(E\,|\,C)$ is the probability of 
{\it event $E$}  assuming that {\it event $C$} occurred.\footnote{In other words, in these terms of {\it cause} and {\it effect}, 
$P(E\,|\,C)$ is the probability 
of $C$ to produce $E$, while $P(C\,|\,E)$ is the probability
of $E$ to have been produced by that cause.} 
The Bayes' theorem tells us how to go from 
$P(E \,|\,C)$ to $P(C\,|\,E)$,  i.e. the probability that 
$C$ is in fact the cause of the event $E$. 
This is indeed the problem we have to solve in order to evaluate the effectiveness of our biometric device.

In its text-book version Bayes' theorem states:
\begin{eqnarray}
P(C\,|\,E) &=& \frac{P(E\,|\,C)\cdot P(C)}
                      {P(E\,|\,C)\cdot P(C) +
                       P(E\,|\,\overline{C})\cdot 
                       P(\overline{C})}\,, 
\end{eqnarray}
where $P(C)$ is the {\it a priori} probability of {\it event $C$}, i.e. the probability of  the occurrence of {\it event $C$} without any information about that of {\it event $E$}.

Without making things too complicate, we can spot the 
too often neglected point:
\begin{center}
{\bf The probability of interest depends on prior probabilities}
\end{center}
In our case it means that 
in order to evaluate the probability of \mbox{LA}, we must have 
a sound estimate of the probability that a person, randomly selected in the population under scrutiny, is not a fake. 
\subsection{A numerical example}
Let us make the example of a population of 1000 people, 
e.g. that of a medium size company, and let us 
assume\footnote{e.g. on the basis of historical or literature data.}
that just a single person could have some reasons to 
act against the company and would be interested 
in selling sensitive data to a competitor. That is 
\begin{eqnarray}
P(\mbox{{\it `unfaithful employee'}}) = P(F) &=& 0.001\,.
\end{eqnarray}
Furthermore, 
let us suppose we have an access control system 
based on biometrics with\footnote{Typical values for fingerprints, by far the most mature biometric technique, are an order of magnitude higher. Iris scanning is alleged to achieve better performance, but it is more expensive and not largely spread.} 
\begin{eqnarray}
\mbox{FMR} &=& 0.001 \\
\mbox{FNMR} &=& 0.001\,.
\end{eqnarray}
that is
\begin{eqnarray}
P(\overline{A}\,|\,F) = 1-P(A\,|\,F) = 1-\mbox{FMR} &=& 0.999\\
P(\overline{F}) =  1-P(F) &=& 0.999\\
P(\overline{A}\,|\,\overline{F}) = \mbox{FNMR} &=& 0.001
\end{eqnarray}
Applying Bayes' theorem, we find that the probability $P(F\,|\,\overline{A})$ that the user is a fake if he/she is not allowed to enter the system, 
given by\,\footnote{It might be convenient to write the expression of 
NLA and LA in terms of the quantities of interest 
[and indicating $P(F)$ simply as $F$], getting thus
\begin{eqnarray*} 
\mbox{NLA} &=& \frac{(1-\mbox{FMR})\cdot F}
                      {(1-\mbox{FMR})\cdot F +
                       \mbox{FNMR}\cdot (1-F)}\\
\mbox{LA} &=& \frac{(1-\mbox{FNMR})\cdot (1-F)}
                      {(1-\mbox{FNMR})\cdot (1-F) +
                       \mbox{FMR}\cdot F}\,.
\end{eqnarray*}
}
\begin{eqnarray}
\mbox{NLA} = P(F\,|\,\overline{A}) &=& \frac{P(\overline{A}\,|\,F)\cdot P(F)}
                      {P(\overline{A}\,|\,F)\cdot P(F) +
                       P(\overline{A}\,|\,\overline{F})\cdot 
                       P(\overline{F})}\,,
\end{eqnarray}
is just 50\%!! In other words, half of accesses denied should have been allowed instead.
\\
The following table shows how the final probability 
changes varying the {\it priors} 
by orders of magnitude\\ 
\begin{center}
\begin{tabular}{c|c}
$P(F)$ & $P(F\,|\,\overline{A}\,)$ \\ 
\hline 
$0.001$ & $0.50$ \\
$0.01$ & $0.91$ \\ 
$0.1$ & $0.99$
\end{tabular} 
\end{center}
Let us also see what happens if we increase both FMR and FNMR 
by one order of magnitude, i.e. $\mbox{FMR} = \mbox{FNMR} = 0.01$.
The previous table becomes now
\begin{center}
\begin{tabular}{c|c}
$P(F)$ & $P(F\,|\,\overline{A}\,)$ \\ 
\hline 
$0.001$ & $0.09$ \\
$0.01$ & $0.50$ \\ 
$0.1$ & $0.92$
\end{tabular} 
\end{center}
The conclusion is evident: the physical and logical environment where
the system is planned to be installed are key issues to be taken 
into account before deciding if adopting a biometric authentication
system and what should be its features and parameters.\footnote{For an elementary important application of Bayes's theorem in the much more serious field of cancer prevention see Ref. \cite{Paulos}}

\section{Other biometric issues}
\subsection{Non-universality}
Biometric is related to physical or behavioural features. Before implementing such a system we have to ask ourselves: ``What if some features are not owned by the whole population?'' Voice recognition does not work with speech impaired people, there are people with unusable fingerprints, gait is not suitable for people on wheelchairs, face could be not available for religious reasons (see below). Alternative authentication mechanisms shall be in place for the persons that can not be {\it measured}.\footnote{In fact the usage of the so called multimodal biometrics (multibiometrics), i.e. the usage of systems based on multiple biometric features has been proposed and analyzed by a theoretical standpoint. Nonetheless it does not seem to have reached the status of a mature product, likely because of the major costs that make multimodal systems not yet suitable for general purpose installations.}
\subsection{Biometric spoofing}
Even though biometrics refers to features strictly individual, they can be spoofed. In other words, someone can steal our data and our identity. Indeed, most of our biometric data are publicly available. We leave our fingerprints everywhere, our voice can be recorded any time, our face is always to the front.

Years ago, the German hacker group Chaos Computer Club  was able to take the fingerprints of the German Ministry of Interior Affairs W. Schauble, publishing 4000 copies as a thin plastic foil,\footnote{\href{http://www.theregister.co.uk/2008/03/30/german_interior_minister_fingerprint_appropriated/}{http://www.theregister.co.uk/2008/03/30/german\_interior\_minister\_fingerprint\_}
\href{http://www.theregister.co.uk/2008/03/30/german_interior_minister_fingerprint_appropriated/}{appropriated/}.
It has to be pointed out that the more sophisticated sensors are able to distinguish a real living finger from a plastic copy by sensing temperature, sweat or heart pulse.} ready to be pasted on fingers of anyone 
wishing to be a Ministry for one day.

While if a password or a digital certificate gets compromised we could change it, revoking it and issuing a new one, once a biometric datum is compromised the remediation could be quite hard.\footnote{We can not enroll more than ten fingers!}
Therefore, how to manage the possible compromising of data shall be analyzed in advance, well before deploying the system, and the remediation steps already designed and operational when the system starts running 
in earnest. 
\subsection{Data protection and identity thefts}
Another aspect often neglected in the CSI-style\footnote{The popular crime drama television series (\href{http://en.wikipedia.org/wiki/CSI:_Crime_Scene_Investigation}{http://en.wikipedia.org/wiki/CSI:\_Crime\_} \href{http://en.wikipedia.org/wiki/CSI:_Crime_Scene_Investigation}{Scene\_Investigation}) portraying a bit unrealistically the forensics investigations.} dissemination of biometrics regards the algorithm transforming the biometric features in raw (analog) data and then in the final byte string (the so-called {\it template}) stored in the device itself or in a centralized database.

\underline{First problem}: Biometrics is still ruled by proprietary solutions, kept secret and protected by patents. That bars an independent evaluation of the device performances and of its real capability to sift the fake users from the legitimate ones.\footnote{The generally deprecated security by obscurity.} 
We have to take for granted \mbox{FMR} and \mbox{FNMR} self-declared by the supplier. 

\underline{Second problem}: Left aside \mbox{FMR} and \mbox{FNMR}, it is quite hard to prove, apart from the claims of the supplier, if the devices are really effective in securing biometric data against an unauthorized usage.\footnote{What happens if the template binary string is presented to the matching algorithm without the physical presence of the unaware owner of the original biometric data? How is this risk avoided?} Even when using a  match-on-card system,\footnote{\url{http://www.matchoncard.com/}} where the template is stored in the device itself, maybe ensuring that no biometric database exists, quite often nothing is said about the security features protecting the template itself.\footnote{Compare that situation to what is required (in many EU countries) for the Secure Signature Creation Devices (typically smart cards) to be used for the qualified electronic signature (\cite{EUDIR, cwa}). They shall be formally certified against a suitable Protection Profile \cite{isocc} by an officially accredited National Certification Body.} 
 
\underline{Third problem}: the template is obviously a digest of the complete measured analog data set. Combining this point with the probabilistic nature of biometrics implies that, in a quite large population, the probability of collision becomes no more irrelevant. So, the probability that a biometric datum is attached to the wrong person becomes not vanishing, like in the famous case of  Brandon Mayfield,\footnote{\url{http://en.wikipedia.org/wiki/Brandon_Mayfield}} that was prosecuted for the Madrid terrorist attack on the basis of a DNA sampling by the U.S. Federal Bureau of Investigation, blindly trusting its technological tools. In this matter Spanish police, using more traditional methods and combining several bodies of evidence\footnote{See \cite{Colombo} again, comparing how the probability updates with a single evidence and with several pieces of evidence, each one less convincing.} put the real terrorists in jail.

And what if someone steals the database where the templates are stored? What is the impact on the end users and what risks are they undertaking?
Nowadays the identity theft is a really serious threat; making easier a villain puts the hands on our own features does not look a brilliant idea. 

\subsection{Changing the scope}
On the basis of what has been discussed in the previous sections, a further important risk arises by carelessly extending the biometric authentication model to a wider and different context compared with the original which the system was designed for. A previously acceptable values of \mbox{FMR} and \mbox{FNMR}, thought for a small strongly characterized population set, can be quite unsuitable in a totally different environment. If, for instance, we put in place an authentication scheme based on face recognition, it is very different to identify or authenticate an employee in a set of some hundreds at most and use the same system in order to identify a person out of millions, maybe in the middle of a crowd.\footnote{Even human brains, well suited in recognizing faces, can be very uncertain about the identification, if a person is met in an unusual place, as it is quite a common experience.}  
So, when we decide to implement a biometric system, the objectives and the borders of the system shall be clearly defined since the very beginning of the project design phase. In fact, a practice that should be followed when building any system.
\subsection{Social and cultural issues, privacy}
Besides security and technical issues, when implementing a biometric system we need to take into account also the social and cultural aspects. 

While there is almost no cultural hurdle to enter a code on a keyboard, using our own body can be at odds with traditional culture, religious prescriptions and similar.
A typical example is the face. Sometimes women wear a veil for religious reasons; men may have bushy beards for religious or traditional habits, hiding most of the face distinctive features. A face recognition system could not work properly with them, because we can not force people to adapt their habits and principles to our technology; likewise it could not work properly with a person enrolled with the spectacles when replaced with the contact lens because such change can dramatically alter the
correspondence between the stored template and the real person under scrutiny. 
But also other systems have their drawbacks: religious beliefs can be at odds with DNA sampling, fingerprints make us feel uncomfortable and a bit criminal. 
All these issues can make hard using a specific biometric system on large scale, while relatively easy to set up in a limited context, with a reduced set of people, more or less willing to give away a digital part of themselves. 

It's not by chance that several Data Protection Authorities (at least in the EU) remarked the proportionality principle between the scope of an authentication scheme and the means to implement it. By itself, this principle is obvious to anyone working with risk analysis, in Information Security or in Insurance, but 
it is not spread enough in other sectors, where too many are prey of astonishing promises peddled by smart sellers.
\section{Conclusions}
In order to fight the spread of the identity frauds we need to add new instruments to our security toolbox and biometrics can surely be one of them. Nevertheless biometrics can be part 
of a wider authentication system only after a careful analysis of the scope and of the specific usage context. Such analysis shall take into account the alternative security measures and the risks of abuse and infringement of 
freedom and dignity. 

Any uncritical acceptance of proposals based on technology only, even if supported by some interesting benchmarks, shall be avoided. If the benchmarks themselves are missing, then we must be wary of the proposer's seriousness. 

All these consideration are especially important when a company supplies 
identification/authenti\-cation services. 
Introducing a biometric system means the company takes charge of all the risks quoted above. 

Too often, this risk engagement is not carefully evaluated as the present paper suggests.

\section*{Acknowledgement}
I would like to thank Giulio D'Agostini, Arturo Salvatici and Daniela Vescio for the accurate reading and the valuable suggestions. I account for  any further mistake.

\end{document}